%% file: Proceedings.tex
\documentclass{PoS}

\usepackage{graphicx}                              
\usepackage{amsmath,amsfonts}
\graphicspath{{./pics/}}                       

\usepackage{epsfig}
\usepackage{amssymb}
\usepackage{amsfonts}
\usepackage{bbm}

\newcommand{\preprintline}{\newline
\vskip -5.2cm
\rightline{\parbox{4cm}{\large\rm  HU-EP-09/52\\ DESY 09-179}}
\vspace{4.2cm}} 

\title{Upper and lower Higgs boson mass bounds from a lattice Higgs-Yukawa model with dynamical overlap fermions \preprintline}
 
\ShortTitle{Upper Higgs boson mass bounds}
 
\author{\speaker{Philipp Gerhold}\\
        Institut f\"ur Physik, Humboldt-Universit\"at zu Berlin, 12489 Berlin, Germany\\
        E-mail: \email{gerhold@physik.hu-berlin.de}}

\author{Karl Jansen\\
        DESY, 15738 Zeuthen, Germany\\
        E-mail: \email{Karl.Jansen@desy.de}}

\include{Commands}

\abstract{We study a lattice Higgs-Yukawa model emulating the same Higgs-fermion coupling structure as 
in the Higgs sector of the electroweak Standard Model, in particular, obeying a Ginsparg-Wilson version 
of the underlying $\mbox{SU}(2)_L\times \mbox{U}(1)_Y$ symmetry, being a global symmetry here due to the
neglection of gauge fields in this model. In this paper we present our results on the 
cutoff-dependent upper Higgs boson mass bound at several selected values of the cutoff parameter $\Lambda$.}

\FullConference{The XXVII International Symposium on Lattice Field Theory\\
		 July 26-31, 2009\\
		 Peking University, Beijing, China}
 
\begin{document}

\section{Introduction}
\label{sec:Introduction}

With the existing evidence for the triviality of the Higgs sector of the electroweak Standard Model, 
rendering the removal of the cutoff $\Lambda$ from the theory impossible, physical quantities
in this sector will, in general, depend on the cutoff. Though this restriction strongly 
limits the predictive power of any calculation performed in the Higgs sector, it opens up
the possibility of drawing conclusions on the energy scale $\Lambda$ at which new physics has
to set in, once, for example, the Higgs boson mass has been determined experimentally.

The main target of lattice studies of the Higgs-Yukawa sector of the electroweak Standard Model 
has therefore been the non-perturbative determination of the cutoff-dependence
of the upper and lower bounds of the Higgs boson mass~\cite{Holland:2003jr,Holland:2004sd} 
as well as its decay properties. There are two main developments which 
warrant the reconsideration of these questions. First, with the advent of the LHC, 
we are to expect that properties of the Standard Model Higgs boson, such as 
the mass and the decay width, will be revealed experimentally. Second, there 
is, in contrast to the situation of earlier investigations of lattice 
Higgs-Yukawa  models~\cite{Smit:1989tz,Shigemitsu:1991tc,Golterman:1990nx,book:Jersak}, 
a consistent formulation of a Higgs-Yukawa model with an exact 
lattice chiral symmetry~\cite{Luscher:1998pq} based on the Ginsparg-Wilson 
relation~\cite{Ginsparg:1981bj}, which allows to emulate the chiral character of the 
Higgs-fermion coupling structure of the Standard Model on the lattice
while lifting the unwanted fermion doublers at the same time.

Since the question for the lower Higgs boson mass bound as well as the phase structure of the
underlying model has already been addressed in \Refs{Gerhold:2009ub,Gerhold:2007yb,Gerhold:2007gx,Fodor:2007fn},
we will here focus only on the discussion of our results concerning the cutoff-dependent upper Higgs 
boson mass bound.

\section{The $\mbox{SU}(2)_L\times \mbox{U}(1)_Y$ lattice Higgs-Yukawa model}
\label{sec:model}

The model we consider here is a four-dimensional lattice Higgs-Yukawa model
with a global $SU(2)_L \times U(1)_Y$ symmetry~\cite{Luscher:1998pq}, aiming 
at the implementation of the chiral Higgs-fermion coupling structure of the 
pure Higgs-Yukawa sector of the Standard Model reading
\beq
\label{eq:StandardModelYuakwaCouplingStructure}
L_Y = y_b \left(\bar t, \bar b \right)_L \varphi b_R 
+y_t \left(\bar t, \bar b \right)_L \tilde\varphi t_R  + c.c,
\eeq
with $y_{t,b}$ denoting the top and bottom Yukawa coupling constants.
Here we have restricted ourselves to the consideration of the top-bottom
doublet $(t,b)$ interacting with the complex scalar doublet $\varphi$ 
($\tilde \varphi = i\tau_2\varphi^*,\, \tau_i:\, \mbox{Pauli-matrices}$), 
since the dynamics of $\varphi$, containing the Higgs mode, is dominated 
by the coupling to the heaviest fermions. For the same reason we also 
neglect any gauge fields in this approach. 

The fields considered in this model are thus the aforementioned doublet $\varphi$ as well as 
$N_f$ top-bottom doublets represented by eight-component spinors 
$\bar\psi^{(i)}\equiv (\bar t^{(i)}, \bar b^{(i)})$, $i=1,...,N_f$.
The chiral character of the targeted coupling structure~(\ref{eq:StandardModelYuakwaCouplingStructure}) 
can then be preserved on the lattice by constructing the fermionic action $S_F$ on the basis of the Neuberger 
overlap operator~\cite{Neuberger:1998wv} according to
\bea
\label{eq:DefYukawaCouplingTerm}
S_F = \sumFL\,
\bar\psi^{(i)}\, \fermiMat\, \psi^{(i)}, 
&\quad&
\fermiMat = \D + 
P_+ \phi^\dagger \fhs{1mm}\mbox{diag}\left(y_t, y_b\right) \hat P_+
+ P_- \fhs{1mm}\mbox{diag}\left(y_t, y_b\right) \phi \hat P_-,
\eea
where the scalar field $\varphi_x$ has been rewritten as a quaternionic, $2 \times 2$ matrix 
$\phi^\dagger_x = (\tilde \varphi_x, \varphi_x)$, with $x$ denoting the site index of the $L_s^3\times L_t$-lattice.
The left- and right-handed projection operators $P_{\pm}$ and the modified projectors $\hat P_{\pm}$
are given as
\bea
P_\pm = \frac{1 \pm \gamma_5}{2}, \quad &
\hat P_\pm = \frac{1 \pm \hat \gamma_5}{2}, \quad &
\hat\gamma_5 = \gamma_5 \left(\ID - \frac{1}{\rho} \D \right),
\eea
with $\rho$ being the radius of the circle of eigenvalues in the complex plane of the free
Neuberger overlap operator~\cite{Neuberger:1998wv}.

This action now obeys an exact global $\mbox{SU}(2)_L\times \mbox{U}(1)_Y$ 
lattice chiral symmetry. For $\Omega_L\in \mbox{SU}(2)$ and $\epsilon\in \re$ the action is invariant under the transformation
\bea
\label{eq:ChiralSymmetryTrafo1}
\psi\rightarrow  U_Y \hat P_+ \psi + U_Y\Omega_L \hat P_- \psi,
&\quad&
\bar\psi\rightarrow  \bar\psi P_+ \Omega_L^\dagger U_{Y}^\dagger + \bar\psi P_- U^\dagger_{Y}, \\
\label{eq:ChiralSymmetryTrafo2}
\phi \rightarrow  U_Y  \phi \Omega_L^\dagger,
&\quad&
\phi^\dagger \rightarrow \Omega_L \phi^\dagger U_Y^\dagger
\eea
with the compact notation $U_{Y} \equiv \exp(i\epsilon Y)$ denoting the respective representations of the 
global hypercharge symmetry group $U(1)_Y$ for the respective field it is acting on. In the continuum 
limit \eqs{eq:ChiralSymmetryTrafo1}{eq:ChiralSymmetryTrafo2} eventually recover the (here global) 
continuum $\mbox{SU}(2)_L\times \mbox{U}(1)_Y$ chiral symmetry.

Finally, the purely bosonic part $S_\varphi$ of the total lattice action $S=S_F+S_\varphi$ is given by the usual 
lattice $\Phi^4$-action
\bea
\label{eq:ContinuumPhiAction}
S_\varphi &=& \sum_{x} \left\{\frac{1}{2}\nabla^f_\mu\varphi_x^{\dagger} \nabla^f_\mu\varphi_x
+ \frac{1}{2}m_0^2\varphi_x^{\dagger}\varphi_x + \lambda\left(\varphi_x^{\dagger}\varphi_x\right)^2   \right\},
\eea
with the bare mass $m_0$, the forward derivative operator $\nabla^f_\mu$ in direction $\mu$, and the bare quartic coupling constant $\lambda$.

\section{Upper Higgs boson mass bounds}
\label{sec:UpperBounds}

In the following the aim will be to determine the largest Higgs boson mass attainable in the considered
Higgs-Yukawa model for a given cutoff $\Lambda$, while being in consistency with phenomenology. Here, the 
later requirement is translated into three matching conditions fixing the vacuum expectation value $v$ of 
the scalar field $\varphi$ as well as the top and bottom quark masses, according to 
\bea
\label{eq:treeLevelTopMass}
246\, \mbox{GeV} = \frac{v_r}{a} \equiv \frac{v}{\sqrt{Z_G}\cdot a}, \quad&
\Lambda = a^{-1}, \quad&
y_{t,b} = \frac{m_{t,b}}{v_r},
\eea
where we restrict ourselves to the mass degenerate case with $m_t/a=m_b/a=\GEV{175}$ in order to guarantee
the fermion determinant $\det(\fermiMat)$ to be real. As a starting point, we simply use the
tree-level relations in \eq{eq:treeLevelTopMass} to fix the bare Yukawa coupling constants. The actually resulting
fermion masses have explicitly been computed on the lattice, which would eventually allow for a more precise tuning
of the Yukawa coupling constants beyond the tree-level relation in some follow-up studies. However, though not 
explicitly demonstrated in this paper, it is found that the Yukawa coupling constants determined by
\eq{eq:treeLevelTopMass} already reproduce the targeted fermion masses with a deviation smaller than $\proz{2}$
in the here considered parameter setups.
The cutoff parameter $\Lambda$ can then (non-uniquely) be defined as the inverse lattice spacing $a^{-1}$, which
is obtained by matching the lattice result on the renormalized vacuum expectation value $v_r=v/\sqrt{Z_G}$ with its
phenomenological value. The underlying Goldstone renormalization constant $Z_G$ is given as
\bea
Z^{-1}_G = \derive{}{p_c^2} \left[\tilde G^{c}_G(p_c^2)\right]^{-1}\Big|_{p_c^2 = -m^2_G},\quad&
\tilde G_G(p) = \frac{1}{3}\sum\limits_{\alpha=1}^3 \langle \tilde g^\alpha_p \tilde g^\alpha_{-p}\rangle, \quad &
\tilde G_H(p) = \langle \tilde h_p \tilde h_{-p}\rangle, 
\eea
with $\tilde G_{H,G}(p)$ denoting the lattice propagators of the Higgs and Goldstone modes in 
momentum space, respectively. For the details of how the aforementioned modes $\tilde h_p$, $\tilde g^\alpha_p$,
and the vacuum expectation value $v$ are extracted from the scalar field $\varphi$, the interested reader is referred 
to \Ref{Gerhold:2009ub}. The Goldstone mass $m_{Gp}$ is then given by the pole of the Goldstone propagator, according to
\bea
\label{eq:DefOfPropagatorMinkMass}
\left[\tilde G_G^{c}(p_c^2)\right]^{-1}\Big|_{p_c^2 = -m^2_{Gp}} = 0, &\quad& 
\RE\left(\left[\tilde G_H^{c}(p_c^2)\right]^{-1}\right)\Big|_{p_c^2 = -m^2_{Hp}} = 0.
\eea
Following the proposition in \Ref{Luscher:1988uq} the Higgs boson mass, on the other hand, is obtained here as the zero 
of the real part of the inverse Higgs propagator, being very close to the actual pole of the propagator~\cite{Luscher:1988uq}
while being numerically much better accessible. 

\includeFigDouble{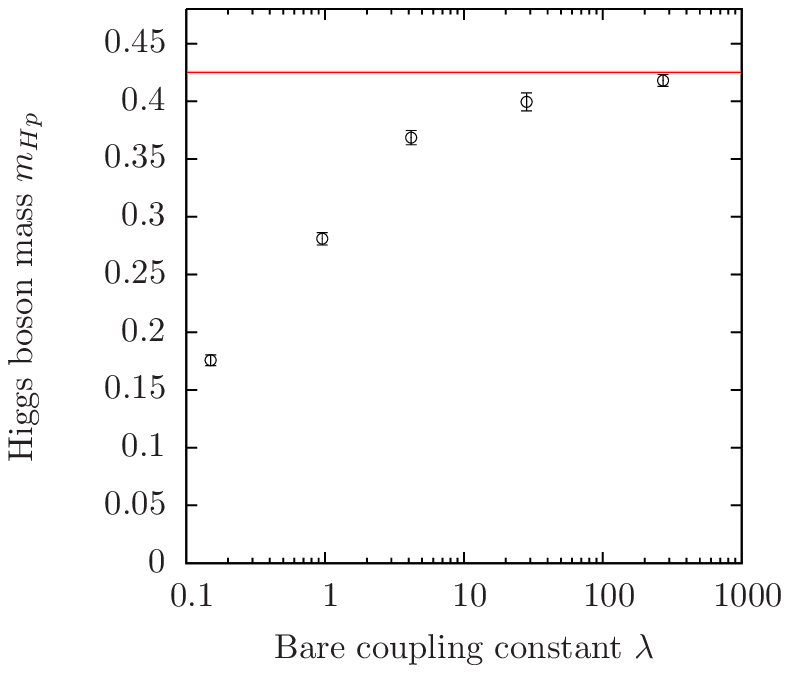}{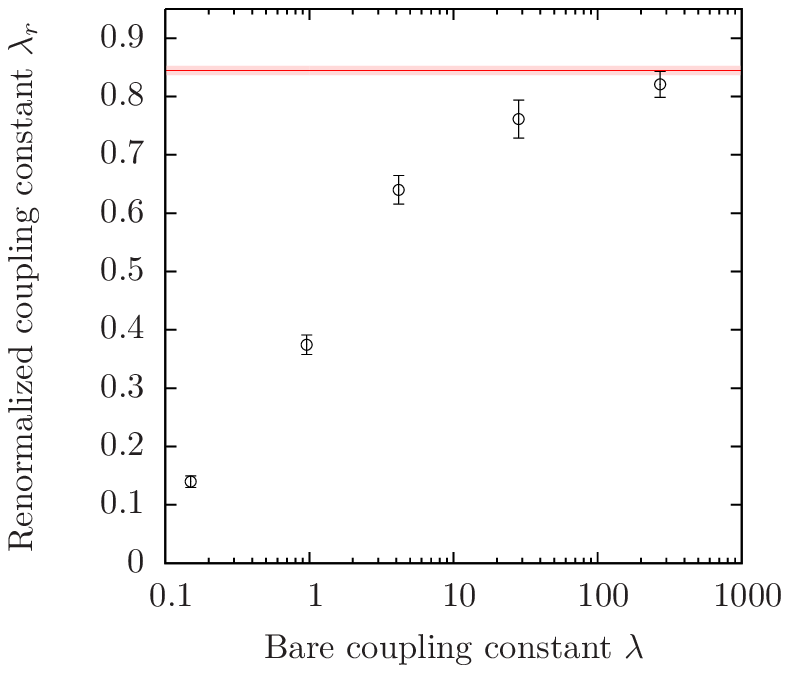}
{fig:StrongCoulingHiggsMassDepOnQuartCoup}
{The Higgs boson mass $m_{Hp}$ and the renormalized quartic coupling constant $\lambda_r$ are shown versus the 
bare coupling constant $\lambda$ in panels (a) and (b), respectively. These results have been obtained in 
direct Monte-Carlo calculations on a \lattice{16}{32} with $N_f=1$. The cutoff parameter $\Lambda$ was intended to be kept
constant, while the actually obtained values of $\Lambda$ fluctuate here between $\GEV{1504}$ and
$\GEV{1549}$. The horizontal lines depict the corresponding results at $\lambda=\infty$, and
the highlighted bands mark the associated statistical uncertainties.
}
{Dependence of the Higgs boson mass $m_{Hp}$ and the renormalized quartic coupling constant $\lambda_r$ on the 
bare quartic coupling constant $\lambda$.}{3}{1}

For clarification it is remarked that $\tilde G^c_{H,G}(p_c)$ denote analytical continuations of the lattice
propagators $\tilde G_{H,G}(p)$, since the latter are only defined at the discrete set of lattice momenta $p$. These 
analytical continuations have been obtained here by fitting the lattice propagators with fit functions derived from
renormalized perturbation theory. 
As discussed in \Ref{Luscher:1988uq} the renormalized quartic coupling constant $\lambda_r$ can then be
defined as
\beq
\label{eq:DefOfRenQuartCoupling}
\lambda_r = \frac{m^2_{Hp}-m^2_{Gp}}{8v_r^2}.
\eeq
 
From perturbation theory one would expect the largest Higgs boson mass to be observed at infinite bare quartic 
coupling constant, \ie $\lambda=\infty$. In \fig{fig:StrongCoulingHiggsMassDepOnQuartCoup} it is explicitly checked
that the renormalized quartic coupling constant $\lambda_r$ as well as the Higgs boson mass $m_{Hp}$ itself are indeed
monotonic functions of the bare parameter $\lambda$, converging to their respective maximum at $\lambda=\infty$,
as expected. For the purpose of determining the upper Higgs boson mass bound, the setting $\lambda=\infty$ is
therefore adapted in the following.

For the eventual determination of the cutoff-dependent Higgs boson mass bound $\upBound$ several series of 
Monte-Carlo calculations have been performed at different values of $\Lambda$ and on different lattice volumes
to finally allow for an infinite volume extrapolation. In order to tame finite volume effects as well as
cutoff effects, we demand here as a minimal requirement that all particle masses $\hat m=m_{Hp}, m_{t}, m_b$ 
in lattice units fulfill 
\bea
\label{eq:RequirementsForLatMass}
\hat m < 0.5& \quad \mbox{and} \quad & \hat m\cdot L_{s,t}>2,
\eea
at least on the largest investigated lattice volumes. Assuming the Higgs boson mass to be below $\GEV{700}$ this
allows to reach energy scales between $\GEV{1400}$ and $\GEV{2800}$ on a \latticeX{32}{32}{.}

In the following we use $N_f=1$, $L_t=32$, and $L_s=12,16,20,24,32$ while the bare mass parameter $m_0$ is tuned
to cover the aforementioned interval of accessible energy scales. In addition, corresponding lattice calculations
have also been performed in the pure $\Phi^4$-theory, \ie with $y_t=y_b=0$, in order to estimate the strength of the 
fermionic contributions to the upper mass bound $\upBound$. The obtained finite volume lattice data are presented 
in \fig{fig:UpperHiggsCorrelatorBoundFiniteVol}.

\includeFigDouble{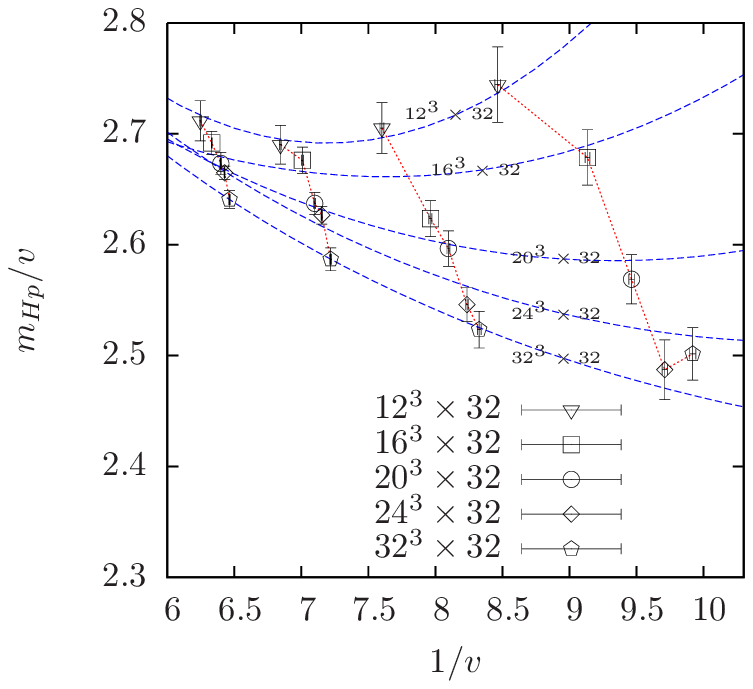}{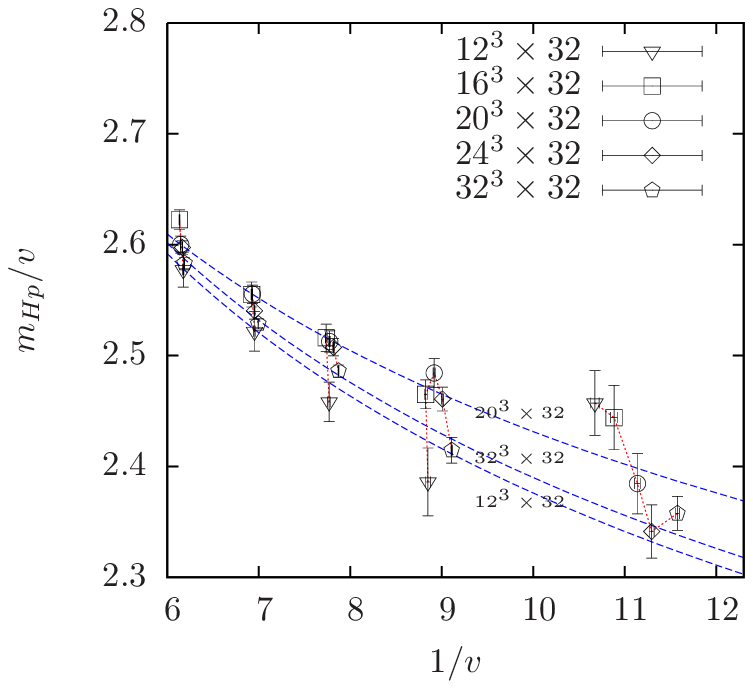}
{fig:UpperHiggsCorrelatorBoundFiniteVol}
{The Higgs propagator mass $m_{Hp}$ is presented in units of the vacuum expectation value $v$ 
versus $1/v$. Those Monte-Carlo results with identical parameter sets differing 
only in the underlying lattice volume are connected via dotted lines to illustrate the effects of the 
finite volume. The dashed curves depict the fits of the lattice results according to the finite size fit approach
in \eq{eq:RenormHiggsMassFitFormula} as explained in the main text. Panel (a) refers to the full Higgs-Yukawa model,
while panel (b) shows the corresponding results of the pure $\Phi^4$-theory.
}
{Dependence of the Higgs propagator mass on $1/v$ at infinite bare quartic coupling constant.}{3}{2}

In order to understand the strong finite volume effects observed in \fig{fig:UpperHiggsCorrelatorBoundFiniteVol}a
we consider here the constraint effective potential $U[\breve v]$. In \Ref{Gerhold:2007yb} it has been derived 
for the degenerate case $y_t=y_b$ in the large $N_f$-limit with $\lambda\propto N_f^{-1}$ and $y_{t,b}\propto N_f^{-1/2}$. It then reads
\bea
\label{eq:EffPot}
U[\breve v] =
\frac{1}{2} m_0^2 \breve v^2 + \lambda \breve v^4 +  U_{F}[\breve v],
&\quad&
U_{F}[\breve v] =
\frac{-4N_f}{L_s^3\cdot L_t}\cdot \sum\limits_{p} \log\left|\nu^+(p) + y_t \breve v \left(1-\frac{1}{2\rho}\nu^+(p)\right) 
\right|^2,\quad\quad
\eea
where $U_F[\breve v]$ denotes here the fermionic contribution and $\nu^+(p)$ is the eigenvalue of the free overlap
Dirac operator with non-negative imaginary part associated to the lattice momentum $p$.

\includeFigDouble{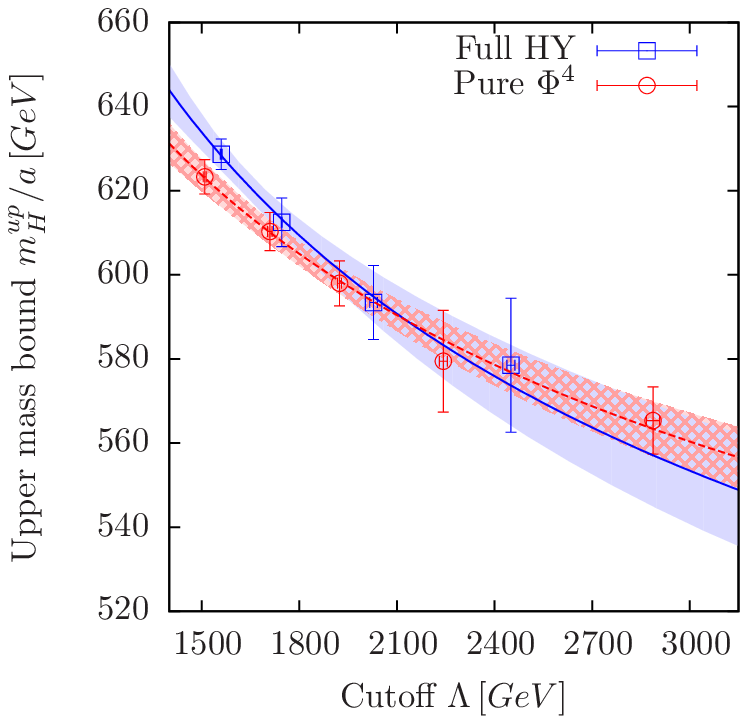}{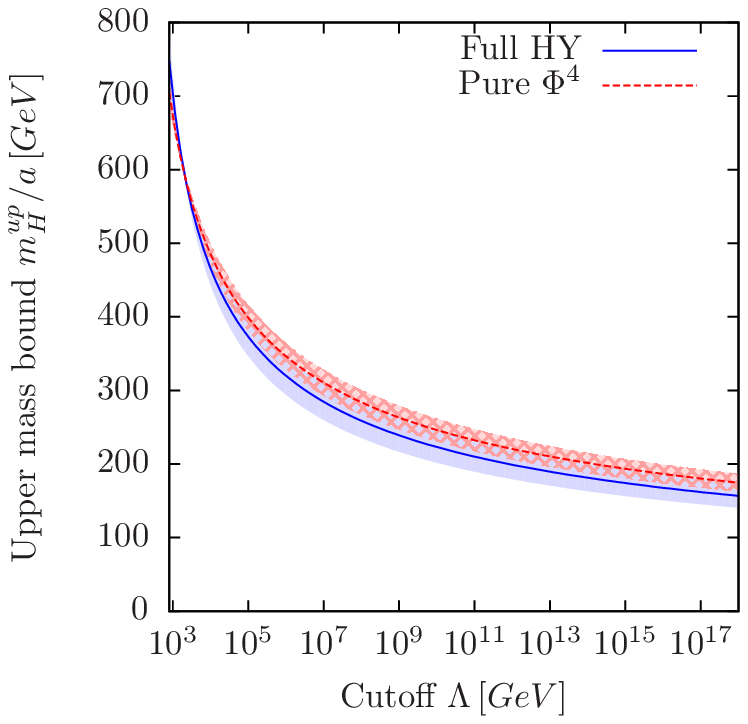}
{fig:UpperMassBoundFinalResult}
{The cutoff dependence of the upper Higgs boson mass bound is presented in panel (a) as obtained from the infinite volume extrapolation results
of the data in \fig{fig:UpperHiggsCorrelatorBoundFiniteVol}. The dashed and solid curves are fits of the data arising from the full Higgs-Yukawa 
model and the pure $\Phi^4$-theory, respectively, with the analytically expected cutoff dependence in \eq{eq:StrongCouplingLambdaScalingBeaviourMass}.
Panel (b) shows the latter fit curves extrapolated to larger values of the cutoff $\Lambda$. In both panels the highlighted bands 
reflect the uncertainty of the respective fit curves.
}
{Cutoff dependence of the upper Higgs boson mass bound.}{3}{2}
 
An estimate $m_{He}$ of the Higgs boson mass can then be obtained from the curvature of the effective potential 
at its minimum, given by the vacuum expectation value $v$, yielding then 
\beq
\label{eq:PropMassFromU}
m^2_{He} = 8 \lambda v^2 - \frac{1}{v} \frac{\mbox{d}}{\mbox{d}\breve v} U_F[\breve v] \Big|_{\breve v = v}
+ \frac{\mbox{d}^2}{\mbox{d}\breve v^2} U_F[\breve v]\Big|_{\breve v = v},
\eeq
which predicts the numerical results on $m_{Hp}$ well in the weak coupling regime~\cite{Gerhold:2009ub}, \ie for $\lambda\ll 1$.
In this case, however, we consider the setting $\lambda=\infty$, rendering thus \eq{eq:PropMassFromU} inapplicable
here. Replacing the bare parameter $\lambda$ with its renormalized counterpart $\lambda_r$, which is well justified
at the considered order in $\lambda$, and exploiting also the expected functional form of the cutoff-dependence 
of $\lambda_r$, which has been derived in \Ref{Luscher:1988uq} according to
\bea
\label{eq:StrongCouplingLambdaScalingBeaviourLamCoupling}
\lambda_r &=& A_\lambda \cdot \left[\log(\Lambda^2/\mu^2) + B_\lambda \right]^{-1},
\eea
where double-logarithmic terms have been neglected, $\mu$ denotes some unspecified scale, and 
$A_{\lambda}\equiv A_{\lambda}(\mu)$, $B_{\lambda}\equiv B_{\lambda}(\mu)$ are constants,
one directly arrives at the expression 
\bea
\label{eq:RenormHiggsMassFitFormula}
m_{He}^2 &=& \frac{8v^2 A_\lambda}{\log(v^{-2}) + B_\lambda}  -\frac{1}{v} \derive{}{\breve v} U_F[\breve v]\Bigg|_{\breve v = v} + \derive{^2}{\breve v^2} 
U_F[\breve v]\Bigg|_{\breve v = v}, 
\eea
which has been used to fit the finite volume lattice data in \fig{fig:UpperHiggsCorrelatorBoundFiniteVol} with the free
fit parameters $A_\lambda$, $B_\lambda$. From the good agreement between the analytical fit curves and the numerical
data one learns that the finite size effects are well understandable already with the simple ansatz given in \eq{eq:RenormHiggsMassFitFormula}.
In particular, the finite size effects in \fig{fig:UpperHiggsCorrelatorBoundFiniteVol}a, which are much stronger than in 
\fig{fig:UpperHiggsCorrelatorBoundFiniteVol}b, can mainly be ascribed to the fermionic contributions. This is also what one would 
have expected, since the top quark is the lightest particle in the scenario considered here.

After having performed an infinite volume extrapolation of the finite size lattice data, the obtained 
results of that extrapolation are finally presented in \fig{fig:UpperMassBoundFinalResult}a. These numerical
data are moreover fitted with the analytically expected functional form of the cutoff-dependence of the Higgs boson mass, derived
in \Ref{Luscher:1988uq} according to
\bea
\label{eq:StrongCouplingLambdaScalingBeaviourMass}
\frac{m_{Hp}}{a} &=& A_m \cdot \left[\log(\Lambda^2/\mu^2) + B_m \right]^{-1/2}, 
\eea
with $A_m\equiv A_{m}(\mu)$, $B_m\equiv B_{m}(\mu)$ denoting the free fit parameters and $\mu$ being again some unspecified scale here. One 
learns from this presentation that the expected logarithmic decline of the Higgs boson mass with increasing cutoff parameter $\Lambda$ can 
very well be resolved. The fermionic contribution to the upper Higgs boson mass bound, however, can not clearly be identified with the 
statistics available here. Finally, it is tempting to extend the fit curves to very large values of $\Lambda$. This has been done in 
\fig{fig:UpperMassBoundFinalResult}b. One finds that the resulting cutoff-dependent upper Higgs boson mass bound would reach a value around 
$\GEV{160}$ at the Planck scale, which is in consistency with earlier perturbative studies within the given uncertainties.
  
\section*{Acknowledgments}
We thank J. Kallarackal for discussions and M. M\"uller-Preussker for his continuous support.
We are grateful to the "Deutsche Telekom Stiftung" for supporting this study by providing a Ph.D. scholarship for
PG. We further acknowledge the support of the DFG through the DFG-project {\it Mu932/4-1}.
The numerical computations have been performed on the {\it HP XC4000 System}
at the {Scientific Supercomputing Center Karlsruhe} and on the
{\it SGI system HLRN-II} at the {HLRN Supercomputing Service Berlin-Hannover}.

\nocite{*}
\bibliographystyle{unsrtOwnNoTitles}  
\bibliography{Proceedings} 
 
\end{document}

%% file: Commands.tex
\newcommand{\vs}{\vspace}
\newcommand{\hs}{\hspace}

\newcommand{\bdm}{\begin{displaymath}}
\newcommand{\edm}{\end{displaymath}}
\newcommand{\beq}{\begin{equation}}
\newcommand{\eeq}{\end{equation}}
\newcommand{\bea}{\begin{eqnarray}}
\newcommand{\eea}{\end{eqnarray}}
\newcommand{\bit}{\begin{itemize}}
\newcommand{\eit}{\end{itemize}}
\newcommand{\bc}{\begin{center}}
\newcommand{\ec}{\end{center}}
\newcommand{\re}{\relax{\rm I\kern-.18em R}}
\newcommand{\ID}{\mathbbm{1}}

\newcommand{\fhs}[1]{\mbox{\hs{#1}}}
\newcommand{\ie}{{\it i.e. }}
\newcommand{\Dov}{{\cal D}^{(ov)}}
\newcommand{\D}{\Dov}

\newcommand{\sumFL}{\sum\limits_{i=1}^{N_f}}

\newcommand{\fermiMat}{{\cal M}}

\newcommand{\latVol}[2]{\ifnum #1=#2 $#1^4$ \else $#1^3\times #2$\fi}
\newcommand{\lattice}[2]{\ifnum #1=#2 $#1^4$-lattice \else $#1^3\times #2$-lattice\fi}
\newcommand{\latticeX}[3]{\ifnum #1=#2 $#1^4$-lattice#3 \else $#1^3\times #2$-lattice#3\fi}
\newcommand{\lattices}[2]{\ifnum #1=#2 $#1^4$-lattice \else $#1^3\times #2$-lattices\fi}
\newcommand{\eq}[1]{Eq.~(\ref{#1})}
\newcommand{\eqs}[2]{Eq.~(\ref{#1}-\ref{#2})}

\newcommand{\fig}[1]{Fig.~\ref{#1}}

\newcommand{\Ref}[1]{Ref.~\cite{#1}}
\newcommand{\Refs}[1]{Refs.~\cite{#1}}
\newcommand{\GEV}[1]{#1\,\mbox{GeV}}

\newcommand{\upBound}{m_{H}^{up}(\Lambda)}
\newcommand{\proz}[1]{#1\,\%}
\newcommand{\RE}{\mbox{Re}}

\newcommand{\derive}[2]{\frac{\mbox{d}#1}{\mbox{d}#2}}

\newcommand{\includeFigDouble}[7]{
\vs{-#6mm}
\bc
\begin{figure}[htb]
\centering
\begin{tabular}{cc}
\includegraphics[width=0.48\textwidth]{#1}
&
\includegraphics[width=0.48\textwidth]{#2}
\\
\hs{4mm}(a) & \hs{8mm}(b)  \\
\end{tabular}
\caption[#5]{#4}
\label{#3}
\vs{-2mm}
\end{figure}
\ec
\vs{-6mm}
\vs{-#7mm}
}